\documentclass[sigconf, nonacm]{acmart}

\usepackage{dsfont}

\newcommand{\empiredb}{{\tt EmpireDB}}

\begin{document}

\title{\empiredb: Data System to Accelerate Computational Sciences}

\author{Daniel Alabi}
\affiliation{}

\author{Eugene Wu}
\affiliation{}

\begin{abstract}

The emerging discipline of \textit{Computational Science} is concerned with using
computers to simulate or solve scientific problems. These problems span
the natural, political, and social sciences. The discipline has
exploded over the past decade due to the emergence of larger amounts of
observational data and large-scale simulations that were previously unavailable or
unfeasible. However, there are still significant challenges with managing the
large amounts of data and simulations.

The database management systems
community has always been at the forefront of the development of the theory and
practice of techniques for formalizing and actualizing 
systems that access or query large datasets.
In this paper, we present \textbf{\empiredb}, 
a vision for a data management system
to accelerate computational sciences. In addition, we
identify challenges and opportunities for the database
community to further the fledgling field of computational sciences.
Finally, we present preliminary evidence showing that the
optimized components in
\empiredb~could lead to improvements in performance compared to
contemporary implementations.

\end{abstract}

\maketitle

\section{Introduction}

Early on his career,
one of the authors was a Database Kernel Engineer on the Distributed Systems team at MongoDB, Inc. The team was responsible for designing and implementing protocols for executing database queries on data that is distributed across multiple machines. The query plan was automatically decided based on several factors (including read/write throughput, data locality, and data distribution).
A shard key (i.e., a single indexed field or multiple fields) was used to distribute data into multiple chunks on different servers. 
The range of shard key options could lead to different data distributions. The
shard key choice is especially important throughout the lifetime of executing different queries on the same data. As a result, domain knowledge of the data distribution and the lifetime of possible queries could be important in the query plan execution. In specific scientific fields (e.g., materials discovery and quantum physics), the data generated can be stored in a flat view. But this view does not take advantage of the data generating process to eliminate redundancies, thus resulting in costly materializations. 
\textit{What if we had a way to design query, execution, and storage plans using knowledge of the scientific domain?} Clearly, this could lead to significant improvements in storage and computation costs.

Computational Science (or Scientific Computing) is an emerging discipline that essentially uses computers to simulate or solve scientific problems (whether in social sciences or natural sciences).
Unfortunately, there are not much data and query models that are well
suited to take advantage of the domain knowledge needed
for computational sciences.
The input of domain knowledge is critical to computational science. Database Theory has been crucial to the design and use of database management systems, providing the SQL (Structured Query Language) interface, the relational model and calculus, and related abstractions~\citep{RG02}. 
Can we apply a similar methodology to the design of database abstractions (such as data and query models) for scientific modeling? 
\textit{We assert that we can!}

Recently, Governor Hochul of the state of New York unveiled
a private and public partnership \textit{Empire AI}, a consortium of
leading research institutions and universities in the state of
New York~\citep{EmpireAI} . The mission of
Empire AI is to bring ``together AI researchers, scientists, entrepreneurs, philanthropists and others'' to make advances in artificial intelligence. The
governor noted that ``the initiative will be funded by over \$400 million in public and private investment.'' One of the focus areas of the collaboration
is to accelerate science by leveraging artificial intelligence (AI)
advances. Insights gained from AI are only accessible when the data can be
stored and retrieved efficiently. However, many scientific datasets are so 
large that it is only feasible to store a small fraction of the
data~\citep{osti_1807223}. We need data systems (and accompanying query and
data models) that can account for 
the storage and approximation of scientific insights. This
is especially important in high-stakes domains (like healthcare
and conservation) where data quality is paramount~\citep{SKHAPA21}.

\begin{figure}
  \centering
  \includegraphics[width=0.95\linewidth]{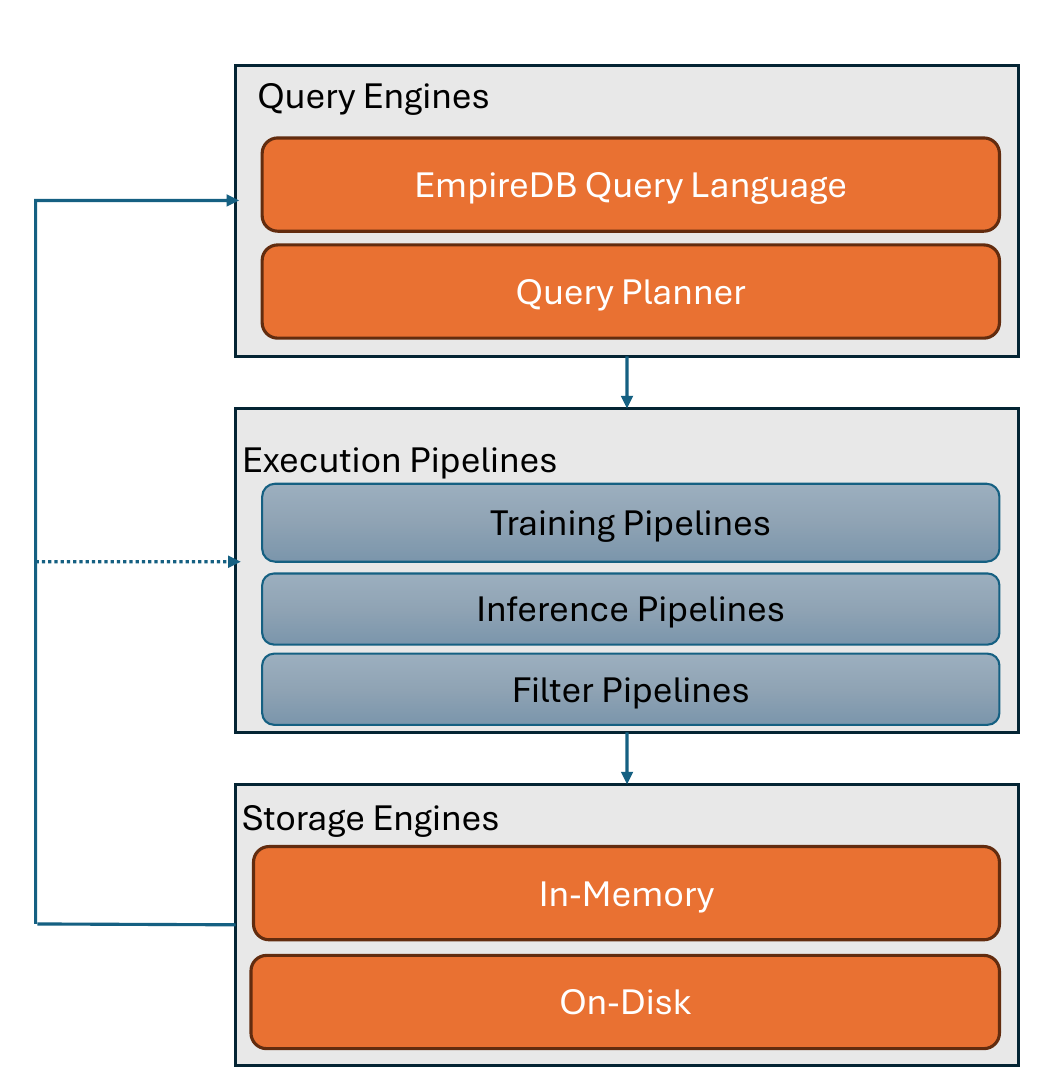}
  \caption{An illustration of system components in \empiredb.}
  \label{fig:empire_db}
\end{figure}

\paragraph{\textbf{Why are existing systems insufficient?}}
As has been previously noted in the database literature,
scientific applications are not served by contemporary relational database systems.
This has led to database components such as SciQL~\citep{zhang2011sciql}, which
provides a language interface where arrays are ``first-class citizens.''
However, in such previous work,
the language interface is not integrated with execution and storage components and thus
falls short of providing an end-to-end vision for computational sciences. In our work,
we provide examples (e.g., in materials discovery) where knowledge of the allowed
approximation errors needs to be: (1) specified via a query language and translated
into a query plan; and (2) passed into subsequent layers for execution of the
training, inference, and filter stages; and (3) stored in memory or on-disk
depending on the required number of active learning stages.
Also, privacy is an important requirement when dealing with
data collected from human subjects~\citep{CSS, huang2023saibot}. 
However, domain knowledge for a particular
application needs to be passed across different layers of a system to gain
scientific insights. 
For example, for the U.S. Census Bureau's
use of privacy protection~\citep{doi:10.1126/sciadv.adl2524} in the 2020 release,
the noisy statistics had to be post-processed for usability sake. e.g.,
the released noisy counts had to be above 0. This post-processing was
a result of the particular use-case for social and political science uses,
with implications for redistricting.
\empiredb~would ensure that domain requirements are easily specified and carried over
across different layers of the system.
In fields ranging from genomics to environmental science, database systems could
facilitate the handling of datasets that are often too large and complex for traditional methods~\citep{osti_1807223}. However, extracting value from context-rich,
heterogeneous sources remains challenging for traditional database systems~\citep{10488724}.
Instead, we could design a system such that the training, inference, and filtering
pipelines (see Figure~\ref{fig:empire_db}) allow cost-based and rule-based
optimization across the entire system~\citep{1677500}.
These could be integrated with existing
system components (e.g., MauveDB, MLLib) that allow
for adapting data processing pipelines to optimize for iterative data flows~\citep{10.1145/1142473.1142483, 10.5555/2946645.2946679, MaiWABHM24}.
In addition, components from multi-dimensional array databases 
(e.g., SciDB~\citep{10.1145/1807167.1807271}) can be incorporated into
\empiredb.

Now let us consider a specific research area:
the quantum many-body physics subfield concerns exploring physical properties of many interacting quantum particles. The interactions between the particles has information that is encoded in some wave function of the entire complex system. Storing and accounting for all interactions becomes infeasible quickly as the dimension of the system scales exponentially with the number of particles. 
\textit{Is there a way to take advantage of query and data
models when performing such complex simulations?}
The recent the GNoME system~\citep{MBSACC23} helps with
materials discovery. However, the system falls short of ways to specify
the allowed levels of approximation in order to discover stable
materials~\citep{SRFKHMMGCMKJBPZC23}.

\textbf{In this work, we introduce our vision for \empiredb, a \textit{novel}
database management system that is
designed to enable and expand computational sciences.}
In Section~\ref{sec:empiredbdesign},
we introduce the general architecture of the system, as well as the
algorithmic optimizations it comes with. Then in Section~\ref{sec:empiredbgnome},
we specialize the framework for a specific use-case in materials
discovery, showcasing the gaps that \empiredb~are  meant to fill.
For the remainder of the paper, we expand on the vision of the \empiredb~system
components and specializations to specific training and inference models.
We also present preliminary experimental results 
that show that the use
of \empiredb~could result in better use of models trained for scientific
purposes.
In the future, the \empiredb~project will be evaluated on the following success criteria:
(i) clearer incorporation of approximation guarantees (via a query model) 
throughout
the lifetime of computational science tasks;
(ii) faster execution of training and inference for scientific findings;
(iii) higher-quality and more accurate models;
(iv) more interpretable models that can be shared and communicated more
easily;
(v) more modular and transferable designs for other tasks in computational
science.

\section{\empiredb: Database System for Computational Sciences}
\label{sec:empiredb}

\subsection{Algorithms and Systems Designs}
\label{sec:empiredbdesign}

Figure~\ref{fig:empire_db} shows the architecture of the \empiredb~system. Scientific computing tasks often rely on approximation methods to
define and evaluate hypotheses. We envision a system where the
tolerance of the approximation can be specified via a query language
(the \empiredb~query language) which would be communicated to other parts of the
system, especially during the training and inference stages.
The training and inference pipelines will be equipped with algorithms
(with provable approximation guarantees) that will be used to generate and
evaluate scientific hypotheses.
Often, scientific inference requires \textit{active learning} stages
(e.g., the GNoME system~\citep{MBSACC23, SRFKHMMGCMKJBPZC23}) so we specify
ways for the storage components to be linked back to the query and execution
pipelines.
\empiredb~has three main components: the query engine, execution pipelines,
and storage engines. Below, we expand on the functionalities of each component:

\subsubsection{Query Engine}

The execution pipeline is where the core operations---training and inference---occur. However, these
pipelines need information about what scientific inference methods should be run,
what approximation tolerance is allowed during inference and training,
and what integrity constraints must be satisfied during computation.
The query engine will be equipped with a language and a planner.

The relational data model was introduced in the 1970s by Codd~\citep{RG02}.
Since then it has permeated every aspect of data management systems.
The data model is arguably the basis of all related fields of data management.
However, in order to incorporate trends in the use of data, it has been
incorporated into specific languages (e.g., the SQL language) and has
evolved to include \textit{integrity constraints} (ICs) that help 
manage the integrity of data flows and storage within a system. In a similar
fashion, with the explosion of data in scientific computing, new
constraints (e.g., for symmetry-aware learning used in the GNoME 
system~\citep{MBSACC23, SRFKHMMGCMKJBPZC23}) must be specified for 
the latest use cases.
The \textbf{query language} will be an extension of the SQL (Structured Query Language) language to include
commands that account for approximation tolerance and constraint-based
learning.
The query language will provide a rich set of syntactic and semantic rules that allow for the expression of complex queries and data manipulations in a readable and efficient manner for scientific computing~\citep{zhang2011sciql}. 

The \textbf{query planner} (also known as the query optimizer) is a critical component of a database management system that determines the most efficient way to execute a given query. For any given scientific task, there could be many methods that could be applied (in the execution pipelines).
It will analyze different possible plans for executing a query and select the one with the lowest cost in terms of resources such as CPU time, memory usage, and disk I/O~\citep{markl2003leo}. 
The query planner can take into account the database schema, the database statistics, and the current state of the system to make its decision. By optimizing the execution plan, the query planner plays a pivotal role in enhancing the performance and responsiveness of the overall
system, especially for such complex queries involving large-scale simulations.

\subsubsection{Execution Pipelines}

The execution pipelines are responsible for training, maintaining, optimizing,
and querying
(machine learning or statistical) models for scientific inference.
These pipelines consist of a series of data preprocessing, feature extraction, model training, and validation steps designed to transform raw data into a model that can make accurate predictions or decisions. 
Each model will provide query answers within a (provable)
tolerance level. There are numerous works in the database literature
that use approximation for developing perceptual models or graph models
(e.g.,~\citep{KerstenIML11, Alabi016, alabi2023degree}).
Training pipelines are automated to ensure consistency and efficiency in model development and to enable the tuning of model parameters. 
For example, a suite of 
Graph Neural Network (GNN) models (with varying layers and/or
input representations)
can be trained in the \textbf{training pipelines}.
The integrity constraints will be enforced during training.

\textbf{Inference pipelines} will
apply trained (statistical or machine learning) models to new data for the purpose of making predictions or decisions. These pipelines handle the loading of the model, preprocessing of input data to match the model requirements, running the model to generate predictions, and post-processing the results if necessary. Inference pipelines are designed for high performance and scalability, enabling real-time or near-real-time decision making.
For example, see recent work on database systems optimized for scaling
queries over video data~\cite{Kang17}.

All models that do not meet the integrity constraints
or approximation tolerance levels will
be filtered out via the \textbf{filter pipeline}.
These pipelines will be implemented as sequences of operations designed to process and refine data models based on specific criteria. They can also
be used to selectively exclude or include data, transform data formats, and apply various processing steps to clean, normalize, or enrich data before it is stored or further analyzed. 
Filter pipelines are essential for maintaining data quality and relevance, ensuring that only pertinent and accurate information is retained for analysis or storage. They enable efficient data management by automating the identification and removal of irrelevant, duplicate, or erroneous data, thus optimizing storage and improving the performance of queries and analyses.
For example, within the GNoME system, a DFT (Density Functional Theory)
sub-component is used to filter candidate novel materials~\citep{MBSACC23}.

\subsubsection{Storage Engines}

An important part of any database management system (DBMS) are 
\textit{file and access} methods that form the basis of storage and
indexing.
By definition, an \textit{index} is a data structure that organizes
records on disk or in-memory to optimize certain query operations~\citep{RG02}.
There are many kinds of index data structures, including tree-based and
hash-based indexing~\citep{ghosh1977data}.
The components of the storage engine will
take advantage of the data generating process to inform the
choice of the index structure. This is particularly helpful for computational
social science and could enable better data sharing paradigms, a core problem
in the field of computational social science~\citep{CSS}.

The \textbf{in-memory} storage engine is a component of a database system designed to store data in the main memory (RAM) of a computer, rather than on disk. This approach significantly accelerates data access and processing speeds because accessing RAM is orders of magnitude faster than accessing disk storage. 
Some scientific inference systems significantly rely on \textit{active learning}
(e.g., the GNoME system~\citep{MBSACC23, SRFKHMMGCMKJBPZC23}) which involves
generating new labeled models to be used in the next inference stage.
These ``fresh'' labeled data should remain in-memory before being fed back
into the training and inference pipelines.

The \textbf{on-disk} storage engines will store and manage data on non-volatile storage devices such as hard drives or SSDs. These engines are optimized for durability and data integrity for large volumes of data. The on-disk storage engines are capable of handling complex transactions, maintaining data consistency, and ensuring data recovery in case of system failures. 
Once the size of the outputs (from the inference pipelines) starts to exceed the
in-memory limits, on-disk storage can be utilized. 

\subsection{Example Problem and Solution Spaces}
\label{sec:empiredbgnome}

\begin{figure*}
  \centering
  \includegraphics[width=0.75\linewidth]{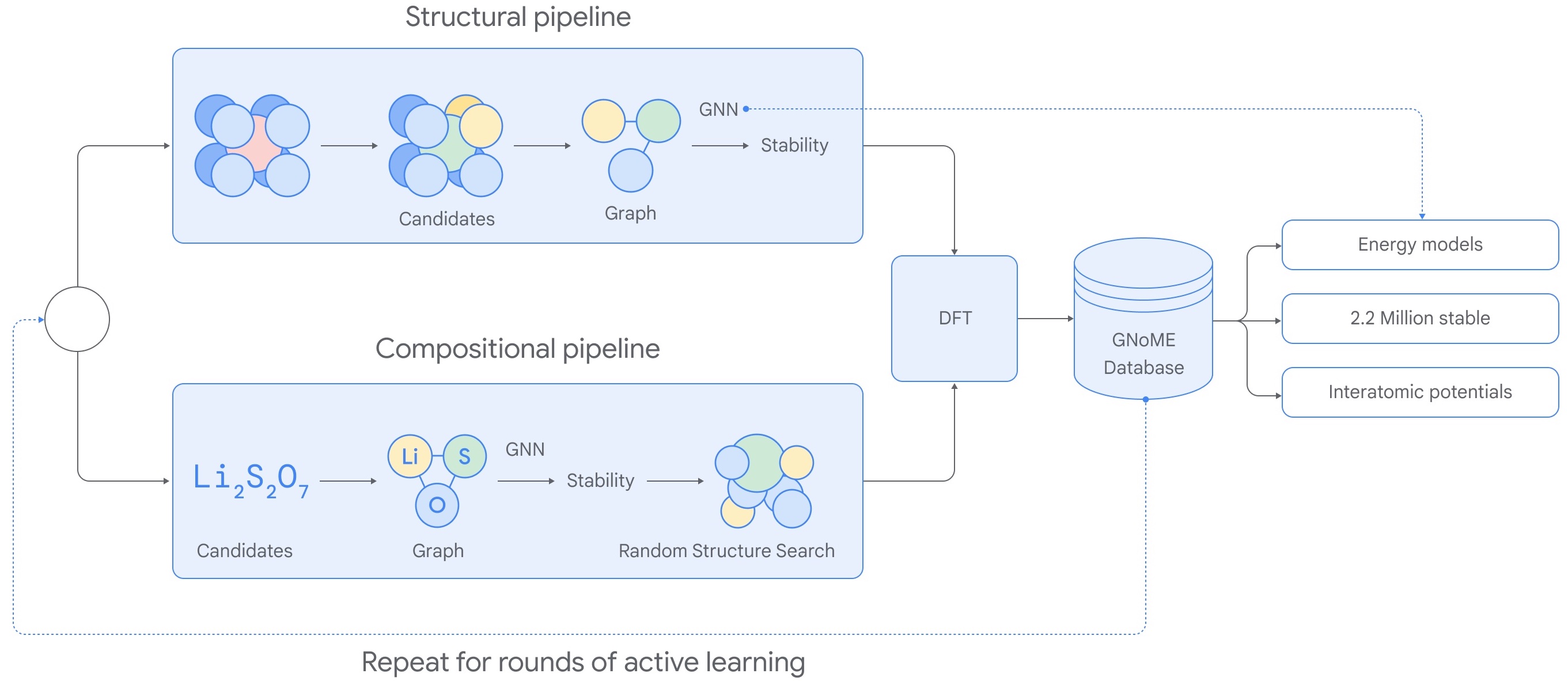}
  \caption{Illustration of the systems architecture of
  GNoME-based materials discovery~\citep{MBSACC23}.}
  \label{fig:gnome_meta}
\end{figure*}

We now discuss some example problems that illustrate
the incorporation of new and revamped database query and data models could
lead to better systems for materials discovery. We note that the problems
require innovations in both algorithms design and systems design.

The field of materials sciences recently has pivoted to substantially
rely on computational methods. While there are many
approaches to simulating materials, most methods on the
microscopic scale involve simulation of quantum mechanics.
Materials consist of atoms; atoms consist of massive, point-like
nuclei (protons and neutrons). 
\textit{The goal is to solve the laws of quantum mechanics for materials
via, for example, calculating ground states of quantum-mechanical systems.}

\paragraph{\textbf{Why is Materials Science via Quantum Mechanics Hard?}}
Take for example, the Fe (Iron) atom. It has 26 electrons which results in
a wave function (mathematical description of the quantum state of an isolated quantum system) of $3\times 26 = 78$ variables. One can attempt to 
store the wave function on a grid. Using a coarse grid of only 10 points
along each direction would require $10^{78}$ numbers! Assuming that
a single precision number stores 4 bytes, then it is \textit{impossible}
to store the description of the wave function.
As a result, approximating the wave function is currently the
only viable approach to solving the laws of quantum mechanics (for materials
sciences, for example).

To synthesize new materials, instead of just
focusing on studying the wave function, we could instead try to 
approximate the space of all possible novel materials in some given class.
This is what GNoME tries to accomplish~\citep{NAP25303}. GNoME finds
candidate materials and then treats the estimation of its energy
as a prediction problem (using information of known materials). Then
after finding
possible candidates, GNoME then uses the Density Functional Theory (DFT)
method to approximate the wave function of a candidate to determine if
the candidate is indeed a stable material. If the material has low-energy,
it is recorded in a database for later use.

\paragraph{\textbf{GNoME for Materials Discovery}}
GNoME uses both
structural properties of previous materials and random search
to generate candidate materials that
are novel and stable.
Using an active learning approach, the GNoME system is able to
identify novel materials and store in a database.
Since the space of all possible novel materials is too large, GNoME
is able to approximate the space as well as the energy of candidate
materials via the use of deep learning.
Specifically, GNoME uses graph neural networks to guide the search
in the space of all novel materials. A GNoME model is a graph neural
network (GNN) that is able to predict the total energy of a material.
The input to the GNNs are one-hot embeddings of
graph representation of materials.

GNoME uses crystal and stability information from an open dataset of
the Materials Project~\citep{JOHCRDCGSCP13}.
GNoME was
trained on a snapshot of the Materials Project from 2021
(containing about 69,000 materials).
The data is used as the basis for all novel discovery.
GNNs are used for model training in both the structural and compositional
pipelines.

The GNoME system (illustrated in
Figure~\ref{fig:gnome_meta})
uses several approximations in its subsystems. The architecture of the 
system for materials discovery was recently adopted by Google DeepMind~\citep{MBSACC23}.
There are four major components of the system:
(1) Structural Pipeline,
(2) Compositional Pipeline,
(3) DFT Component,
(4) GNoME Database.
The database management system (GNoME database) appears to be a
stand-alone component that operates without close integration with
the structural and compositional pipelines. 
We discuss some example opportunities to explore
with the system sub-components:

\subsubsection{Structural Pipeline}

The structural pipeline component of GNoME is used to discover
low-energy or stable materials. It operates by creating 
candidates that are similar in structure to known materials.
It uses ``symmetry-aware partial substitutions'' (SAPS) to introduce changes
to the structure of a crystal while respecting the symmetries
present in the crystal~\citep{togo2018textttspglib}.

For the structural pipelines, the dataset of known materials is
converted into graph structures:
edges are present between atoms that are within
a pre-determined interatomic cutoff. Using 3-6 layers of message passing,
the final layer of the GNNs
output the energy estimate of a candidate material.
Using this information, the candidate material can be fed into the
DFT component.
However, how can one scale the process of introduction of changes to
the crystal structure? What database optimizations could help?

There are a range of database
algorithmic optimizations that potentially could 
be applied for similarity and symmetry-aware search.
To name just one example:
for similarity search and the approximate nearest neighbor
problem, 
LSH (Locality-Sensitive Hashing) techniques can be utilized
~\citep{DIIM04}.
There is an opportunity to integrate such data structures 
and algorithms into the
domain of materials discovery.
For the \empiredb~system, the optimizations will be made accessible
via the execution pipelines.

\subsubsection{Compositional Pipeline}

The compositional pipeline discovers low-energy/stable materials but
works in a different fashion than the structural pipeline. It uses
randomized algorithms to search for stable
structures using chemical formulas of known materials.
Essentially, the algorithms randomly combine elements to create new structures
that obey certain symmetry requirements~\citep{MBSACC23}.
Throughout the process, the GNNs are used to predict the total energy of
a crystal to determine if it is a viable candidate for the next stage.

The compositional pipeline requires certain symmetry requirements.
Is there a way to specify this requirement (via a query language) throughout
the execution of the algorithms?
Recent work explores what a query language for graph-related data might
accomplish~\citep{Geerts23}. The query language interface in \empiredb~will enable
such specification.

Also, from the database community, there are search algorithms that can be
leveraged to speed up the pipeline. For example, B-trees and B+-trees
are used to create indexes within a DBMS. 
Spatial Indexing techniques (e.g., Quad trees) can also be employed in the
pipeline given the high-dimensional nature of materials search
~\citep{RG02}.
These techniques can be incorporated into the storage engine of \empiredb.

\subsubsection{Density Functional Theory (DFT) Calculations}

Both the structural and compositional pipelines are used to discover
novel stable materials. The outputs are then evaluated via
DFT calculations.
The DFT calculations give approximations of physical energies of systems
of molecules. 

When candidate materials are found by either the
structural or compositional pipelines, the energies of the candidates
are evaluated by DFT to verify if the candidate gives a stable (low-energy)
material.
If it does give a stable material, it is then added to the GNoME database
of new materials and the active learning process continues. We believe
the DFT sub-system can be closely integrated with the pipelines to
potentially reduce latency. In particular, the filter pipeline sub-component of
\empiredb~could incorporate the DFT sub-system.
Also, the query language interface of \empiredb~can be extended so that
DFT calculations could be specified via the language.

Closer integration of DFT with the pipelines can be accomplished.
For example, during the random structure search process in the
compositional pipeline, information about the search space size (which
gives measures of uncertainty) from which
a candidate was drawn from can be fed into the DFT. The information about
the search space size could be used to determine if it is worth running
DFT on the candidate material.

The foundational principle of DFT is that the ground state properties of a many-electron system can be uniquely determined by its electron density distribution. This principle significantly simplifies the problem of solving the Schrödinger equation for many-electron systems, which is a daunting task due to the complexity of electron-electron interactions. By converting the many-body problem into a functional of the electron density, DFT makes it possible to study systems of practical size and complexity. The Hohenberg-Kohn theorems formalize this concept, stating firstly that the ground state energy of a system is a unique functional of its electron density, and secondly, that there exists a variational principle for the electron density, 
leading to the ground state energy~\citep{parr1994density, Becke93, PhysRev.140.A1133, PhysRev.136.B864}.

Despite its simplifications, DFT has been remarkably successful in predicting a wide range of physical and chemical properties, such as reaction energies, electronic structures, and magnetic properties, across a vast array of systems from small molecules to bulk materials~\citep{parr1994density, Becke93}. Its ability to provide insights into the electronic behavior of materials underpins its indispensable role in the design of new compounds and materials, including catalysts, semiconductors, and batteries. The continuous development of more accurate exchange-correlation functionals and efficient computational algorithms ensures DFT's ongoing relevance and utility in scientific research.

In conclusion, Density Functional Theory stands as a cornerstone of modern computational science, offering a pragmatic yet powerful approach to understanding the quantum mechanical behavior of matter. Its development and refinement over the years have made it an indispensable tool for researchers across a spectrum of scientific disciplines. As computational power increases and new theoretical advancements are made, DFT is poised to remain at the forefront of materials science, chemistry, and physics, continuing to uncover the mysteries of the material world.
As such, DFT can be applied beyond the GNoME system!
Given the popularity of DFT-based techniques in computational sciences,
we believe that DFT will be an essential component of the filter
pipeline of \empiredb.

\section{Preliminary Experimental Evidence}
\label{sec:experiments}

We present preliminary experimental evidence of how database
optimizations (beyond the query modeling work) could help with
materials science.
There are clearly many dimensions of experiments that one can run to verify
the benefits of \empiredb. We leave the full scale of such experiments to
future work.

For now, we will focus on some modifications to the use of GNN
designs within execution pipelines.
The goal is to classify graphs using structural graph properties
of molecules. The dataset we use is the MUTAG dataset that contains
188 graphs, with an average of 18 nodes and 20 edges.
Table~\ref{tab:train-gnn} shows the training accuracy when using a
system with static number of layers vs. \empiredb~that is able to tune the
number of layers to meet a threshold of accuracy.
In all cases, \empiredb~achieves higher levels of accuracy.

\begin{table}[]
\begin{tabular}{| l | l | l | l |}\hline

Dimension of Hidden Features & 64 & 256  & 512\\\hline
Static \#Layers & 91.19\% & 91.71\% & 93.75\% \\\hline
\empiredb-Tuned & 91.24\% & 94.84\% & 94.85\% \\\hline
\end{tabular}
\caption{Training accuracy for GNN-based models.}
\label{tab:train-gnn}
\end{table}

\section{Conclusion}

In this work, we presented \empiredb, a database system for computational
sciences. We show the relevance of the system in accelerating
materials discovery; using this general framework, \empiredb~can be applied
to other problems that go beyond computational materials science. 
\empiredb~can lead to more effective systems for simulating, illuminating, or solving scientific problems. 
While we have focused on materials discovery, our vision can be extended to
cybersecurity applications as well~\citep{EgressyNBAWA24}, where GNNs are used for
security applications.

The use of database systems in addressing scientific challenges represents a transformative shift in how researchers collect, analyze, and share data~\citep{CSS}. These complex systems serve as the backbone for managing vast quantities of information, enabling scientists to store, retrieve, and manipulate data efficiently. In fields ranging from genomics to environmental science, database systems facilitate the handling of datasets that are often too large and complex for traditional methods~\citep{osti_1807223}.
This capability is particularly valuable in fields like epidemiology and climate science, where understanding complex patterns can lead to breakthroughs in public health and environmental policy~\citep{Knsel2019ApplyingBD}.
By democratizing access to scientific data, database systems play a crucial role in advancing knowledge and innovation in the scientific community~\citep{NAP25303}.
Moreover, database systems play a pivotal role in facilitating collaborative science. In an era where interdisciplinary research is increasingly important, these systems enable seamless sharing and integration of data across different fields and institutions. By standardizing data formats and ensuring interoperability, database systems make it possible for researchers from diverse backgrounds to work together more effectively. 
However, current database systems do not
take advantage of the approximation that can be specified at different layers of the
system nor the active learning required for certain scientific tasks. \empiredb~is
our vision that aims to further the use of database systems for computational
science.

For future work, we will explore the algorithms and systems problems associated
with making \empiredb~scale to larger datasets and to more research
fields.

\bibliographystyle{plainurl}

\bibliography{main}

\begin{thebibliography}{10}

\bibitem{alabi2023degree}
Daniel Alabi and Dimitris Kalimeris.
\newblock Degree distribution identifiability of stochastic kronecker graphs, 2023.
\newblock \href {https://arxiv.org/abs/2310.00171} {\path{arXiv:2310.00171}}.

\bibitem{Alabi016}
Daniel Alabi and Eugene Wu.
\newblock Pfunk-h: approximate query processing using perceptual models.
\newblock In {\em Proceedings of the Workshop on Human-In-the-Loop Data Analytics, HILDA@SIGMOD 2016, San Francisco, CA, USA, June 26 - July 01, 2016}, page~10. {ACM}, 2016.

\bibitem{Becke93}
Axel~D. Becke.
\newblock {Density‐functional thermochemistry. III. The role of exact exchange}.
\newblock {\em The Journal of Chemical Physics}, 98(7):5648--5652, 04 1993.
\newblock \href {https://arxiv.org/abs/https://pubs.aip.org/aip/jcp/article-pdf/98/7/5648/19277469/5648\_1\_online.pdf} {\path{arXiv:https://pubs.aip.org/aip/jcp/article-pdf/98/7/5648/19277469/5648\_1\_online.pdf}}, \href {https://doi.org/10.1063/1.464913} {\path{doi:10.1063/1.464913}}.

\bibitem{10.1145/1807167.1807271}
Paul~G. Brown.
\newblock Overview of scidb: large scale array storage, processing and analysis.
\newblock In {\em Proceedings of the 2010 ACM SIGMOD}, SIGMOD '10, page 963–968, New York, NY, USA, 2010. Association for Computing Machinery.
\newblock \href {https://doi.org/10.1145/1807167.1807271} {\path{doi:10.1145/1807167.1807271}}.

\bibitem{osti_1807223}
Aydin Buluc, Tamara~G. Kolda, Stefan~M. Wild, Mihai Anitescu, Anthony Degennaro, John~D. Jakeman, Chandrika Kamath, Ramakrishnan Kannan, Miles~E. Lopes, Per-Gunnar Martinsson, Kary Myers, Jelani Nelson, Juan Restrepo, C.~Seshadri, Draguna Vrabie, Brendt Wohlberg, Stephen~J. Wright, Chao Yang, and Peter Zwart.
\newblock Randomized algorithms for scientific computing (rasc).
\newblock 7 2021.
\newblock URL: \url{https://www.osti.gov/biblio/1807223}, \href {https://doi.org/10.2172/1807223} {\path{doi:10.2172/1807223}}.

\bibitem{DIIM04}
Mayur Datar, Nicole Immorlica, Piotr Indyk, and Vahab~S. Mirrokni.
\newblock Locality-sensitive hashing scheme based on p-stable distributions.
\newblock In {\em Proceedings of the Twentieth Annual Symposium on Computational Geometry}, SCG '04, page 253–262, New York, NY, USA, 2004. Association for Computing Machinery.
\newblock \href {https://doi.org/10.1145/997817.997857} {\path{doi:10.1145/997817.997857}}.

\bibitem{10.1145/1142473.1142483}
Amol Deshpande and Samuel Madden.
\newblock Mauvedb: supporting model-based user views in database systems.
\newblock In {\em Proceedings of the 2006 ACM SIGMOD}, SIGMOD '06, page 73–84, New York, NY, USA, 2006. Association for Computing Machinery.

\bibitem{EgressyNBAWA24}
B{\'{e}}ni Egressy, Luc von Niederh{\"{a}}usern, Jovan Blanusa, Erik~R. Altman, Roger Wattenhofer, and Kubilay Atasu.
\newblock Provably powerful graph neural networks for directed multigraphs.
\newblock In {\em AAAI}, pages 11838--11846. {AAAI} Press, 2024.

\bibitem{Geerts23}
Floris Geerts.
\newblock A query language perspective on graph learning.
\newblock In {\em Proceedings of the 42nd {ACM} {SIGMOD-SIGACT-SIGAI} Symposium on Principles of Database Systems, {PODS} 2023, Seattle, WA, USA, June 18-23, 2023}, pages 373--379. {ACM}, 2023.

\bibitem{ghosh1977data}
S.P. Ghosh.
\newblock {\em Data Base Organization for Data Management}.
\newblock Computer Science and Applied Mathematics. A Series of Monographs and Textobooks. Academic Press, 1977.
\newblock URL: \url{https://books.google.com/books?id=h-omAAAAMAAJ}.

\bibitem{PhysRev.136.B864}
P.~Hohenberg and W.~Kohn.
\newblock Inhomogeneous electron gas.
\newblock {\em Phys. Rev.}, 136:B864--B871, Nov 1964.
\newblock URL: \url{https://link.aps.org/doi/10.1103/PhysRev.136.B864}, \href {https://doi.org/10.1103/PhysRev.136.B864} {\path{doi:10.1103/PhysRev.136.B864}}.

\bibitem{huang2023saibot}
Zezhou Huang, Jiaxiang Liu, Daniel~Gbenga Alabi, Raul~Castro Fernandez, and Eugene Wu.
\newblock Saibot: A differentially private data search platform.
\newblock {\em Proceedings of the VLDB Endowment}, 16(11):3057--3070, 2023.

\bibitem{JOHCRDCGSCP13}
Anubhav Jain, Shyue~Ping Ong, Geoffroy Hautier, Wei Chen, William~Davidson Richards, Stephen Dacek, Shreyas Cholia, Dan Gunter, David Skinner, Gerbrand Ceder, and Kristin~A. Persson.
\newblock {Commentary: The Materials Project: A materials genome approach to accelerating materials innovation}.
\newblock {\em APL Materials}, 1(1):011002, 07 2013.
\newblock \href {https://doi.org/10.1063/1.4812323} {\path{doi:10.1063/1.4812323}}.

\bibitem{Kang17}
Daniel Kang, John Emmons, Firas Abuzaid, Peter Bailis, and Matei Zaharia.
\newblock Noscope: optimizing neural network queries over video at scale.
\newblock {\em Proc. VLDB Endow.}, 10(11):1586–1597, aug 2017.
\newblock \href {https://doi.org/10.14778/3137628.3137664} {\path{doi:10.14778/3137628.3137664}}.

\bibitem{doi:10.1126/sciadv.adl2524}
Christopher~T. Kenny, Cory McCartan, Shiro Kuriwaki, Tyler Simko, and Kosuke Imai.
\newblock Evaluating bias and noise induced by the u.s. census bureau’s privacy protection methods.
\newblock {\em Science Advances}, 10(18):eadl2524, 2024.
\newblock URL: \url{https://www.science.org/doi/abs/10.1126/sciadv.adl2524}, \href {https://arxiv.org/abs/https://www.science.org/doi/pdf/10.1126/sciadv.adl2524} {\path{arXiv:https://www.science.org/doi/pdf/10.1126/sciadv.adl2524}}, \href {https://doi.org/10.1126/sciadv.adl2524} {\path{doi:10.1126/sciadv.adl2524}}.

\bibitem{KerstenIML11}
Martin~L. Kersten, Stratos Idreos, Stefan Manegold, and Erietta Liarou.
\newblock The researcher's guide to the data deluge: Querying a scientific database in just a few seconds.
\newblock {\em Proc. {VLDB} Endow.}, 4(12):1474--1477, 2011.
\newblock URL: \url{http://www.vldb.org/pvldb/vol4/p1474-kersten.pdf}.

\bibitem{Knsel2019ApplyingBD}
Benedikt Kn{\"u}sel, Marius Zumwald, Christophe Baumberger, Gertrude~Hirsch Hadorn, Erich~M. Fischer, David~N. Bresch, and Reto Knutti.
\newblock Applying big data beyond small problems in climate research.
\newblock {\em Nature Climate Change}, 9:196--202, 2019.
\newblock URL: \url{https://api.semanticscholar.org/CorpusID:91553899}.

\bibitem{PhysRev.140.A1133}
W.~Kohn and L.~J. Sham.
\newblock Self-consistent equations including exchange and correlation effects.
\newblock {\em Phys. Rev.}, 140:A1133--A1138, Nov 1965.
\newblock URL: \url{https://link.aps.org/doi/10.1103/PhysRev.140.A1133}, \href {https://doi.org/10.1103/PhysRev.140.A1133} {\path{doi:10.1103/PhysRev.140.A1133}}.

\bibitem{CSS}
David M.~J. Lazer, Alex Pentland, Duncan~J. Watts, Sinan Aral, Susan Athey, Noshir Contractor, Deen Freelon, Sandra Gonzalez-Bailon, Gary King, Helen Margetts, Alondra Nelson, Matthew~J. Salganik, Markus Strohmaier, Alessandro Vespignani, and Claudia Wagner.
\newblock Computational social science: Obstacles and opportunities.
\newblock {\em Science}, 369(6507):1060--1062, 2020.
\newblock URL: \url{https://science.sciencemag.org/content/369/6507/1060}.

\bibitem{MaiWABHM24}
Anh~L. Mai, Pengyu Wang, Azza Abouzied, Matteo Brucato, Peter~J. Haas, and Alexandra Meliou.
\newblock Scaling package queries to a billion tuples via hierarchical partitioning and customized optimization.
\newblock {\em Proc. {VLDB} Endow.}, 17(5):1146--1158, 2024.

\bibitem{markl2003leo}
Volker Markl, Guy~M Lohman, and Vijayshankar Raman.
\newblock Leo: An autonomic query optimizer for db2.
\newblock {\em IBM Systems Journal}, 42(1):98--106, 2003.

\bibitem{10.5555/2946645.2946679}
Xiangrui Meng, Joseph Bradley, Burak Yavuz, Evan Sparks, Shivaram Venkataraman, Davies Liu, Jeremy Freeman, DB~Tsai, Manish Amde, Sean Owen, Doris Xin, Reynold Xin, Michael~J. Franklin, Reza Zadeh, Matei Zaharia, and Ameet Talwalkar.
\newblock Mllib: machine learning in apache spark.
\newblock {\em J. Mach. Learn. Res.}, 17(1):1235–1241, jan 2016.

\bibitem{MBSACC23}
Amil Merchant, Simon Batzner, Samuel Schoenholz, Muratahan Aykol, Gowoon Cheon, and Ekin Cubuk.
\newblock Scaling deep learning for materials discovery.
\newblock {\em Nature}, 624:1--6, 11 2023.
\newblock \href {https://doi.org/10.1038/s41586-023-06735-9} {\path{doi:10.1038/s41586-023-06735-9}}.

\bibitem{NAP25303}
National~Academies of~Sciences \& Engineering~\& and Medicine.
\newblock {\em Reproducibility and Replicability in Science}.
\newblock The National Academies Press, Washington, DC, 2019.
\newblock URL: \url{https://nap.nationalacademies.org/catalog/25303/reproducibility-and-replicability-in-science}, \href {https://doi.org/10.17226/25303} {\path{doi:10.17226/25303}}.

\bibitem{EmpireAI}
New York Governor~Press Office.
\newblock Governor hochul unveils fifth proposal of 2024 state of the state: Empire ai consortium to make new york the national leader in ai research and innovation.
\newblock \url{https://www.governor.ny.gov/news/governor-hochul-unveils-fifth-proposal-2024-state-state-empire-ai-consortium-make-new-york}.
\newblock Accessed: 2024-02-21.

\bibitem{parr1994density}
R.G. Parr and Y.~Weitao.
\newblock {\em Density-Functional Theory of Atoms and Molecules}.
\newblock International Series of Monographs on Chemistry. Oxford University Press, 1994.
\newblock URL: \url{https://books.google.com/books?id=mGOpScSIwU4C}.

\bibitem{RG02}
Raghu Ramakrishnan and Johannes Gehrke.
\newblock {\em Database Management Systems}.
\newblock McGraw-Hill, Inc., USA, 3 edition, 2002.

\bibitem{SKHAPA21}
Nithya Sambasivan, Shivani Kapania, Hannah Highfill, Diana Akrong, Praveen Paritosh, and Lora~M Aroyo.
\newblock “everyone wants to do the model work, not the data work”: Data cascades in high-stakes ai.
\newblock In {\em Proceedings of the 2021 CHI Conference}, CHI '21, 2021.

\bibitem{10488724}
Viktor Sanca and Anastasia Ailamaki.
\newblock Efficient model-relational data management: Challenges and opportunities.
\newblock {\em IEEE Transactions on Knowledge and Data Engineering}, pages 1--12, 2024.
\newblock \href {https://doi.org/10.1109/TKDE.2024.3384276} {\path{doi:10.1109/TKDE.2024.3384276}}.

\bibitem{SRFKHMMGCMKJBPZC23}
Nathan Szymanski, Bernardus Rendy, Yuxing Fei, Rishi Kumar, Tanjin He, David Milsted, Matthew McDermott, Max Gallant, Ekin Cubuk, Amil Merchant, Haegyeom Kim, Anubhav Jain, Chris Bartel, Kristin Persson, Yan Zeng, and Gerbrand Ceder.
\newblock An autonomous laboratory for the accelerated synthesis of novel materials.
\newblock {\em Nature}, 624:1--6, 11 2023.
\newblock \href {https://doi.org/10.1038/s41586-023-06734-w} {\path{doi:10.1038/s41586-023-06734-w}}.

\bibitem{togo2018textttspglib}
Atsushi Togo and Isao Tanaka.
\newblock $\texttt{Spglib}$: a software library for crystal symmetry search, 2018.
\newblock \href {https://arxiv.org/abs/1808.01590} {\path{arXiv:1808.01590}}.

\bibitem{1677500}
T.F. Wenisch, R.E. Wunderlich, M.~Ferdman, A.~Ailamaki, B.~Falsafi, and J.C. Hoe.
\newblock Simflex: Statistical sampling of computer system simulation.
\newblock {\em IEEE Micro}, 26(4):18--31, 2006.

\bibitem{zhang2011sciql}
Ying Zhang, Martin Kersten, Milena Ivanova, and Niels Nes.
\newblock Sciql: bridging the gap between science and relational dbms.
\newblock In {\em Proceedings of the 15th symposium on international database engineering \& applications}, pages 124--133, 2011.

\end{thebibliography}

\end{document}